# Performance Bounds for Parameter Estimation under Misspecified Models: Fundamental findings and applications

S. Fortunati[1], F. Gini[1], M. S. Greco[1], and C. D. Richmond[2]

[1]Dipartimento di Ingegneria dell'Informazione, University of Pisa, Italy

[2]Arizona State University, School of Electrical, Computer and Energy Engineering, Tempe, USA.

**ABSTRACT**

Inferring information from a set of acquired data is the main objective of any signal processing (SP) method. In particular, the common problem of estimating the value of a *vector* of parameters from a set of noisy measurements is at the core of a plethora of scientific and technological advances in the last decades; for example, wireless communications, radar and sonar, biomedicine, image processing, and seismology, just to name a few.

Developing an estimation algorithm often begins by assuming a statistical model for the measured data, i.e. a probability density function (pdf), which if correct, fully characterizes the behaviour of the collected data/measurements. Experience with real data, however, often exposes the limitations of any assumed data model since modelling errors at some level are always present. Consequently, the true data model and the model assumed to derive the estimation algorithm could differ. When this happens, the model is said to be *mismatched* or *misspecified*. Therefore, understanding the possible performance loss or regret that an estimation algorithm could experience under model misspecification is of crucial importance for any SP practitioner. Further, understanding the limits on the performance of any estimator subject to model misspecification is of practical interest.

Motivated by the widespread and practical need to assess the performance of a "mismatched" estimator, the goal of this paper is to help to bring attention to the main theoretical findings on estimation theory, and in particular on lower bounds under model misspecification, that have been published in the statistical and econometrical literature in the last fifty years. Secondly, some applications are discussed to illustrate the broad range of areas and problems to which this framework extends, and consequently the numerous opportunities available for SP researchers.

**1. INTRODUCTION**

The mathematical basis for a formal theory of statistical inference was presented by Fisher, who introduced the Maximum Likelihood (ML) method along with its main properties [Fis25]. Since then, ML estimation has been widely used in a variety of applications. One of the main reasons for its popularity is its asymptotic efficiency, i.e. its ability to achieve a minimum value of the error variance as the number of available observations goes to infinity or as the noise power decreases to zero. The concept of efficiency is strictly related to the existence of some lower bounds on the performance of any estimator designed for a specific inference task. Such performance bounds, one of which is the celebrated Cramér-Rao Bound (CRB) [Cra46] [Rao45], are of fundamental importance in practical applications since they provide a benchmark of comparison for the performance of any estimator. Specifically, given a particular estimation problem, if the performance of a certain algorithm achieves a relevant performance bound, then no other algorithm can do better. Moreover, evaluating a performance bound is often a prerequisite for any feasibility study. In particular, the availability of a lower bound for the estimation problem at hand makes the SP practitioner aware of the practical impossibility to achieve better estimation accuracy than the one indicated by the bound itself. Another fundamental feature of a performance bound is its ability to capture and reveal the complex dependences amongst the various parameters of interest, thus offering the opportunity to understand more deeply the estimation problem at hand and ultimately to identify an appropriate design choice of parameters and criterion for an estimator [Kay98].

Before describing specific performance bounds, it is worth mentioning that estimation theory explores two different frameworks: one is *deterministic* and one is *Bayesian*. In the classical deterministic approach, the parameters to be estimated are modelled as *deterministic* but *unknown* variables. This implies that no a-priori information is available that would suggest that one outcome is more or less likely than another. In the Bayesian framework instead, the parameters of interest are assumed to be *random variables* and the goal is to estimate their particular realizations. Unlike the classical deterministic approach, the Bayesian approach exploits this random characterization of the unknown parameters by incorporating *a-priori information* about the unknown parameters in the derivation of an estimation algorithm. In particular, the joint pdf of the unknown parameters is assumed known, and therefore can be taken into account in the estimation process through Bayes' theorem [Kay98].

## 1.1 BASICS ABOUT PERFORMANCE BOUNDS

When talking about lower bounds, the first distinction that needs to be made is between *local* (or small-error) bounds and *global* (or large-error) bounds. A bound can be considered a local error bound if its calculation relies exclusively on the behaviour of the pdf of the data at a single point value of the parameter (or perhaps a very small "local" neighbourhood around this point). If the calculation of a bound requires knowledge of the pdf behaviour at multiple (more than one) distinct and well separated (non-local) points, then the bound can be characterized as a global error bound. Local error bounds at best determine the limits of the asymptotics of optimal algorithms like ML, whereas characterization of non-asymptotic performance must somehow take into account the possible influence of parameter values other than the true value.

A bound is said to be "tight" if it reasonably predicts the performance of the ML estimator. In particular, if a bound is only asymptotically tight, then it is reliable only in the presence of high Signal-to-Noise Ratio (SNR) or sufficiently large number of measurements. On the other hand, if a bound is globally tight, then it is a reliable bound for the error covariance of an ML estimator irrespective of the SNR level or of the amount of available data. The deterministic bound that can be regarded as the most general representative of the class of global bounds is the Barankin Bound (BB) [Bar49]. However, due to its generality, the calculation of the BB is not straightforward and it usually does not admit a closed form representation. The most popular local bound is the above-mentioned CRB. Unlike the BB, the CRB is easy to evaluate for many practical problems, but it is reliable only *asymptotically*. In the non-asymptotic region, which is characterized by a low SNR and/or by a low number of measurements, the CRB can be "too optimistic" with respect to the effective error covariance achievable by an estimator [Van13].

The second subdivision of the performance bounds is a direct consequence of the dichotomy between the deterministic and the Bayesian estimation frameworks. In particular, we can identify the class of deterministic lower bounds and the class of Bayesian lower bounds [Van07]. Without any claim of completeness, the class of the deterministic lower bounds includes: the (global) BB [Bar49], and two local bounds, the Bhattacharyya Bound [Bha46] and the CRB [Cra46] [Rao45]. We stress that the most common forms of these bounds, including the CRB, apply only to *unbiased* estimators. Versions of these bounds exist, however, that can be applied to biased estimators whose bias function can be determined. Concerning the Bayesian bounds, they can be divided into two classes [Ren08]: the Ziv-Zakaï family and the Weiss-Weinstein family, to which the Bayesian version of the CRB belongs. The first family is derived by relating mean squared error to the probability of error in a binary hypothesis testing problem, while the derivation of the latter is based on the covariance inequality. For further details on Bayesian Bounds, we refer the reader to the comprehensive book [Van07].

1.2 AN ESTIMATION THEORY UNDER MODEL MISSPECIFICATION: MOTIVATIONS

Regardless of the differences previously discussed, both the classical deterministic estimation theory and the Bayesian framework are based on the implicit assumption that the assumed data model, i.e. the pdf, and the true data model are the same, i.e. the model is *cor-* *rectly specified*. However, much evidence from engineering practice shows that this assumption is often violated, i.e. the assumed model is different from the true one. There are two main reasons for *model misspecification*. The first is the imperfect knowledge of the true data model, which leads to an incorrect specification of the data pdf. On the other hand, there could be cases where perfect knowledge of the true data model is available but, due to an intrinsic computational complexity or to a costly hardware implementation, it is not possible nor convenient to pursue the optimal "matched" estimator. In these cases, one may prefer to derive an estimator by assuming a simpler but misspecified data model, e.g. the Gaussian model. Of course, this suboptimal procedure may lead to some degradation in the overall system performance, but ensures, on the other hand, a simple analytical derivation and real-time hardware implementation of the inference algorithm. It is clear that, in such a misspecified estimation framework, the possibility to assess the impact of the model misspecification on the estimation performance is of fundamental importance to guarantee the reliability of the (mismatched) estimator. Misspecified bounds are then the perfect candidates to fulfil this task: they generalize the classical framework by allowing the assumed and true models to differ, yet they establish performance limits on the estimation error covariance in a way that indicates how the difference between the true and assumed models affects the estimation performance. Having established the main motivations, we can now briefly review the literature on the estimation framework under model misspecification, with a focus on the two classical building blocks, i.e. the ML estimator and the CRB.

1.3 SOME HISTORICAL BACKGROUND

The first fundamental result on the behaviour of the ML estimator under misspecification appeared in the statistical literature in the year 1967 and was provided by Huber [Hub67]. In that paper the consistency and the normality of the ML estimator were proved under very mild regularity conditions. Five years later, Akaike [Aka72] highlighted the link between Huber's findings and the Kullback-Leibler divergence (KLD) [Cov06]. He noted that the convergence point of the ML estimator under model misspecification could be interpreted as the point that minimizes the KLD between the true and the assumed models. In the early 80's, these ideas were further developed by White [Whi82], where the term "Quasi-Maximum Likelihood" (QML) estimator was introduced. Some years later, the second fundamental building block of an estimation theory under model misspecification was established by Vuong in [Vuo86]. Vuong was the first one to derive a generalization of the Cramér-Rao lower bound under misspecified models. Analysis of the Bayesian misspecified estimation problem has been investigated in [Ber66] and [Bun98].

Quite surprisingly and despite the wide variety of potential applications, the SP community has remained largely unaware of these fundamental results. Only recently, this topic has been rediscovered and its applications to well-known SP problems investigated ([Xu04], [Noa07], [Ric13], [Gre14], [Gus14], [Ric15], [Kan15], [Par15], [Ren15], [Fri15], [For16a], [For16b], [For16c], [Ric16], [Men18]). Of course, every SP practitioner was well aware of the misspecification problem, but some approaches commonly used within the SP community to address it differed from some of those proposed in the statistical literature. In particular, the effect of the misspecification has been modelled by adding into the true data model some random quantities, also called *nuisance parameters*, and by transforming the estimation problem at hand into a higher dimensional *hybrid* estimation problem. The performance degradation due to the augmented level of uncertainty generated by the nuisance parameters could be assessed by evaluating the true CRB when possible, the hybrid CRB (see, e.g. [Roc87], [Gin00], [Par08], [Noa09]), or the modified CRB ([And92], [Gin98], [Kba17]). This approach, although reasonable, is application-dependent and not general at all. Other approaches include sensitivity analyses [Fri90], [Van13].

Finally, it could also be of interest to point out here the relationship between misspecified estimation theory and robust (see [Zou12] for an excellent tutorial on robust statistics). As one would expect, these two frameworks share the same motivations, i.e. an imperfect knowledge of the true data model. The aim of robust estimation theory is to develop estimation algorithms capable of achieving *good* performance over a large set of allowable input data mod-els, even if suboptimal under any *nominal* (or true) model. Even though the development of robust estimators are surely of great importance in many SP applications, for some of these the mathematical derivation and consequent implementation may be too involved or too time and hardware intensive. In these cases, as discussed before, one may prefer to apply the classical, non-robust, estimation theory by assuming a simplified, hence misspecified, statistical model for the data.

The first aim of this article is to summarize the most relevant existing works in the statistical literature using a formalism that is more familiar to the SP community. Secondly, we aim to show the potential application of misspecified estimation theory, in both the deterministic and Bayesian contexts, to various classical SP problems.

## 2. DESCRIPTION OF A MISSPECIFIED MODEL PROBLEM

Let $\mathbf{x}_1,\ldots,\mathbf{x}_M$ be a set of *N*-dimensional (generally complex) random vectors, representing the outcome of a measurement process. Let $\mathbf{x}_m \in \mathbb{C}^N$ be a single observation vector with probability density function (pdf) $p_X(\mathbf{x}_m)$ belonging to a possibly *parametric* family, or *model*, $\mathcal{P}$ that characterizes the observed random experiment. As discussed in Sect. 1, in almost all practical applications the *true* pdf $p_X(\mathbf{x}_m)$ either is not perfectly known, or it does not admit a simple derivation or easy implementation of the estimation algorithm. Thus, instead of $p_X(\mathbf{x}_m)$, in the mismatched estimation framework we adopt a different *parametric* pdf, say $f_X(\mathbf{x}_m | \boldsymbol{\theta})$ with $\boldsymbol{\theta} \in \Theta \subset \mathbb{R}^d$, to characterize the statistical behaviour of the data vector $\mathbf{x}_m$. Potential estimation algorithms may be derived from the *misspecified parametric* pdf $f_X(\mathbf{x}_m | \boldsymbol{\theta})$ belonging to a *parametric model* $\mathcal{F}$ and not from the true pdf $p_X(\mathbf{x}_m)$. Moreover, we assume that $f_X(\mathbf{x}_m | \boldsymbol{\theta})$ could differ from $p_X(\mathbf{x}_m)$ for every $\boldsymbol{\theta} \in \Theta$. Since this assumption represents the division between the classical matched and the misspecified parametric estimation theories, some additional comments are warranted. The matched estimation theory requires the existence of at least a parameter vector $\bar{\boldsymbol{\theta}} \in \Theta$ for which the pdf assumed by the SP practitioner is equal to the true one. Mathematically, we can say that the classical matched theory holds true if, for some $\bar{\boldsymbol{\theta}} \in \Theta$, $p_X(\mathbf{x}_m) = f_X(\mathbf{x}_m | \bar{\boldsymbol{\theta}})$, or equivalently if $p_X(\mathbf{x}_m) \in \mathcal{F}$. For example, suppose that the collected data, i.e. the outcomes of a random experiment, are distributed according to a univariate Gaussian distribution with mean value $\bar{\mu}$ and variance $\bar{\sigma}^2$, namely, $x_m \sim p_X(x_m) = \mathcal{N}(\bar{\mu}, \bar{\sigma}^2)$, $m = 1, \ldots, M$. Moreover, suppose that the assumed parametric model for data inference is the Gaussian parametric model, i.e. $\mathcal{F} = \{f_X | f_X(x_m | \boldsymbol{\theta}) = \mathcal{N}(\theta_1, \theta_2) \ \forall \boldsymbol{\theta} \in \mathbb{R} \times \mathbb{R}^+\}$, where $\mathbb{R}^+$ is the set of positive real numbers. This situation clearly represents a matched case, since there exists $\bar{\boldsymbol{\theta}} = (\bar{\mu}, \bar{\sigma}^2) \in \mathbb{R} \times \mathbb{R}^+$ such that $p_X(x_m) = f_X(x_m | \bar{\boldsymbol{\theta}}) = \mathcal{N}(\bar{\mu}, \bar{\sigma}^2)$. Suppose now that the collected data are actually distributed according to a univariate Laplace distribution with location parameter $\bar{\gamma}$ and scale parameter $\bar{\beta}$, i.e. $x_m \sim p_X(x_m) = \mathcal{L}(\bar{\gamma}, \bar{\beta})$. Due to perhaps misleading and incomplete information on the experiment at hand, or due to the need to derive a simple algorithm, we decide to adopt a parametric Gaussian model $\mathcal{F}$ to characterize the collected data. Unlike the previous case, this is obviously a mismatched case, since there does not exist any $\boldsymbol{\theta} = (\theta_1, \theta_2)$ for which the assumed Gaussian model is equal to the true Laplace model.

Many practical examples of model misspecification can be found in the everyday engineering practices. Just to name a few, recent papers have investigated the application of this misspecifed model framework to the Direction-of-Arrival (DoA) estimation problem in sensor arrays ([Ric13],[Ric15],[Kan15]) and MIMO radars [Ren15], to the covariance matrix estimation problem in non-Gaussian disturbance ([Gre14], [For16a], [For16c]), to radar-communication systems coexistence [Ric16], to waveform parameter estimation in the presence of uncertainty in the propagation model [Par15], and to the Time-of-Arrival (ToA) estimation problem for ultra wideband (UWB) signals in the presence of interference [Gus14].

Since the first part of the paper deals with the deterministic misspecified estimation theory, the parameter vector $\boldsymbol{\theta}$ is assumed to be an *unknown* and *deterministic* real vector. The ex-tension to the Bayesian case is discussed in Sect. 5. Suppose that for inference purposes we collect *M independent, identically distributed* (i.i.d.) measurement vectors $\mathbf{x} = \{\mathbf{x}_m\}_{m=1}^{M}$, where $\mathbf{x}_m \sim p_X(\mathbf{x}_m)$. Due to the independence, the *true* joint pdf of the dataset $\mathbf{x}$ can be expressed as the product of the marginal pdf as $p_X(\mathbf{x}) = \prod_{m=1}^{M} p_X(\mathbf{x}_m)$. The *assumed* joint pdf of the dataset is instead $f_X(\mathbf{x}|\boldsymbol{\theta}) = \prod_{m=1}^{M} f_X(\mathbf{x}_m|\boldsymbol{\theta})$.

This misspecified model framework raises two important questions:

- Is it still possible to derive lower bounds on the error covariance of any mismatched estimator of the parameter vector $\boldsymbol{\theta}$?
- How will the classical statistical properties of an estimator, e.g. unbiasedness, consistency and efficiency, change in this misspecified model framework?
- What is the meaningfulness of parameter estimates under extreme cases of mismatch?

The remainder of this paper addresses these fundamental issues.

### 3. THE MISSPECIFIED CRAMÉR-RAO BOUND

This section introduces a version of the CRB accounting for possible model misspecification, i.e. the misspecified Cramér-Rao bound (MCRB), which can be considered a generalization of the usual CRB, which is obtained when the model is correctly specified. We start by providing the required regularity conditions and the notion of unbiasedness for mismatched estimators.

#### 3.1 REGULAR MODELS

As with the classical CRB, in order to guarantee the existence of the MCRB, some regularity conditions on the assumed pdf need to be imposed. Specifically, the assumed parametric model $\mathcal{F}$ has to be *regular* with respect to (w.r.t.) $\mathcal{P}$, i.e. the family to which the true pdf belongs. The complete list of assumptions that $\mathcal{F}$ has to satisfy to be regular w.r.t. $\mathcal{P}$ are given in [Vuo86] and briefly recalled in [For16a]. Most of them are rather technical and facilitate order reversal of integral and derivative operators. Nevertheless, there are two assumptions that need to be discussed here due to their importance in the development of the theory. The first condition that has to be satisfied is:

*A1.* There exists a unique interior point $\boldsymbol{\theta}_0$ of $\Theta$ such that

$$\boldsymbol{\theta}_0 = \arg\min_{\boldsymbol{\theta} \in \Theta} \{-E_p\{\ln f_X(\mathbf{x}_m|\boldsymbol{\theta})\}\} = \arg\min_{\boldsymbol{\theta} \in \Theta}\{D(p_X \| f_{X|\boldsymbol{\theta}})\}, \qquad (1)$$

where $E_p\{\cdot\}$ indicates the expectation operator of a vector- or scalar-valued function w.r.t. the pdf $p_X(\mathbf{x}_m)$ and $D(p_X \| f_{X|\boldsymbol{\theta}}) \triangleq \int \ln(p_X(\mathbf{x}_m)/f_X(\mathbf{x}_m|\boldsymbol{\theta}))p_X(\mathbf{x}_m)d\mathbf{x}_m$ is the KLD [Cov06] between the true and the assumed pdfs. As indicated by eq. (1), $\boldsymbol{\theta}_0$ can be interpreted as the point that minimizes the KLD between $p_X(\mathbf{x}_m)$ and $f_X(\mathbf{x}_m|\boldsymbol{\theta})$ and it is called the *pseudo-true parameter vector* ([Vuo86], [Whi82]).

After having defined the pseudo-true parameter vector $\boldsymbol{\theta}_0$ in A1, let $\mathbf{A}_{\boldsymbol{\theta}_0}$ be the matrix whose entries are defined as:

$$\left[\mathbf{A}_{\boldsymbol{\theta}_0}\right]_{ij} \triangleq \left[E_p\{\nabla_{\boldsymbol{\theta}}\nabla_{\boldsymbol{\theta}}^T \ln f_X(\mathbf{x}_m|\boldsymbol{\theta}_0)\}\right]_{ij} = E_p\left\{\left.\frac{\partial^2}{\partial \theta_i \partial \theta_j}\ln f_X(\mathbf{x}_m|\boldsymbol{\theta})\right|_{\boldsymbol{\theta}=\boldsymbol{\theta}_0}\right\}, \qquad (2)$$

where $\nabla_{\boldsymbol{\theta}} u(\boldsymbol{\theta}_0)$ and $\nabla_{\boldsymbol{\theta}}\nabla_{\boldsymbol{\theta}}^T u(\boldsymbol{\theta}_0)$ indicate respectively the gradient (column) vector and the symmetric Hessian matrix of the scalar function $u$ evaluated at $\boldsymbol{\theta}_0$. The second fundamental condition that must be satisfied by the assumed model $\mathcal{F}$ in order to be regular w.r.t. $\mathcal{P}$ is:

*A2.* The matrix $\mathbf{A}_{\boldsymbol{\theta}_0}$ is non-singular.

The pseudo-true parameter vector $\boldsymbol{\theta}_0$ plays a fundamental role in estimation theory for misspecified models. Roughly speaking, it identifies the pdf $f_X(\mathbf{x}_m|\boldsymbol{\theta}_0)$ in the assumed parametric model $\mathcal{F}$ that is closest, in the KLD sense, to the true model. As the next sections

will clarify, it can be interpreted as the counterpart of the true parameter vector of the classical matched theory. Regarding the matrix $\mathbf{A}_{\boldsymbol{\theta}_0}$, its negative represents a generalization of the classical Fisher Information Matrix (FIM) to the misspecified model framework. In order to clarify this, we first define the matrix $\mathbf{B}_{\boldsymbol{\theta}_0}$ as:

$$\left[\mathbf{B}_{\boldsymbol{\theta}_0}\right]_{ij} \triangleq \left[E_p\left\{\nabla_{\boldsymbol{\theta}} \ln f_X(\mathbf{x}_m | \boldsymbol{\theta}_0) \nabla_{\boldsymbol{\theta}}^T \ln f_X(\mathbf{x}_m | \boldsymbol{\theta}_0)\right\}\right]_{ij}$$
$$= E_p\left\{\frac{\partial \ln f_X(\mathbf{x}_m | \boldsymbol{\theta})}{\partial \theta_i}\bigg|_{\boldsymbol{\theta}=\boldsymbol{\theta}_0} \cdot \frac{\partial \ln f_X(\mathbf{x}_m | \boldsymbol{\theta})}{\partial \theta_j}\bigg|_{\boldsymbol{\theta}=\boldsymbol{\theta}_0}\right\}. \quad (3)$$

As with matrix $\mathbf{A}_{\boldsymbol{\theta}_0}$, we recognize in $\mathbf{B}_{\boldsymbol{\theta}_0}$ the second possible generalization of the FIM. Vuong [Vuo86] showed that if $p_X(\mathbf{x}_m) = f_X(\mathbf{x}_m | \bar{\boldsymbol{\theta}})$ for some $\bar{\boldsymbol{\theta}} \in \Theta$, then $\boldsymbol{\theta}_0 = \bar{\boldsymbol{\theta}}$ and $\mathbf{B}_{\bar{\boldsymbol{\theta}}} = -\mathbf{A}_{\bar{\boldsymbol{\theta}}}$, where $\bar{\boldsymbol{\theta}}$ is the true parameter vector of the classical matched theory. The last equality shows that, under correct model specification, the two expressions of the FIM are equal, as expected [Van13]. This provides evidence of the fact that the misspecified estimation theory is consistent with the classical one. The reader, however, should note the fact that the equality between the pseudo-true parameter vector and the true one does not imply in any way the equality between the true and the assumed pdfs, and consequently between the matrices $\mathbf{B}_{\boldsymbol{\theta}_0}$ and $-\mathbf{A}_{\boldsymbol{\theta}_0}$. After having established the necessary regularity conditions, we can introduce the class of misspecified-unbiased (MS-unbiased) estimators.

3.2 THE MISSPECIFIED (MS) – UNBIASEDNESS PROPERTY

The first generalization of the classical unbiasedness property to mismatched estimators was proposed by Vuong [Vuo86]. Specifically, let $\hat{\boldsymbol{\theta}}(\mathbf{x})$ be an estimator of the pseudo-true parameter vector $\boldsymbol{\theta}_0$, i.e. a function of the $M$ available i.i.d. observation vectors $\mathbf{x} = \{\mathbf{x}_m\}_{m=1}^{M}$, derived under the misspecified parametric model $\mathcal{F}$. Then $\hat{\boldsymbol{\theta}}(\mathbf{x})$ is said to be an MS-unbiased estimator if and only if (iff):

$$E_p\{\hat{\boldsymbol{\theta}}(\mathbf{x})\} = \int \hat{\boldsymbol{\theta}}(\mathbf{x}) p_X(\mathbf{x}) d\mathbf{x} = \boldsymbol{\theta}_0, \quad (4)$$

where $\boldsymbol{\theta}_0$ is the pseudo-true parameter vector defined in eq. (1). The link with the classical matched unbiasedness property is obvious: if the parametric model $\mathcal{F}$ is correctly specified, $\boldsymbol{\theta}_0$ is equal to the vector $\bar{\boldsymbol{\theta}} \in \Theta$ such that $p_X(\mathbf{x}_m) = f_X(\mathbf{x}_m | \bar{\boldsymbol{\theta}})$. Consequently, eq. (4) can be rewritten as $E_p\{\hat{\boldsymbol{\theta}}(\mathbf{x})\} = \int \hat{\boldsymbol{\theta}}(\mathbf{x}) f_X(\mathbf{x} | \bar{\boldsymbol{\theta}}) d\mathbf{x} = \bar{\boldsymbol{\theta}}$, which is the classical definition of the unbiasedenss property. At this point, we are ready to introduce the explicit expression for the MCRB.

3.3 A COVARIANCE INEQUALITY IN THE PRESENCE OF MISSPECIFIED MODELS

In this section, we present the MCRB as introduced by Vuong in his seminal paper [Vuo86]. An alternative derivation was proposed by Richmond and Horowitz in [Ric13] and [Ric15]. A comparison between the derivation given in [Vuo86] and the one proposed in [Ric13] and [Ric15] has been provided in [For16a].

*Theorem 1 [Vuo86]*: Let $\mathcal{F}$ be a misspecified parametric model that is regular w.r.t. $\mathcal{P}$. Let $\hat{\boldsymbol{\theta}}(\mathbf{x})$ be an MS-unbiased estimator derived under the misspecfied model $\mathcal{F}$ from a set of $M$ i.i.d. observation vectors $\mathbf{x} = \{\mathbf{x}_m\}_{m=1}^{M}$. Then, for every possible $p_X(\mathbf{x}_m) \in \mathcal{P}$:

$$\mathbf{C}_p\left(\hat{\boldsymbol{\theta}}(\mathbf{x}), \boldsymbol{\theta}_0\right) \geq \frac{1}{M} \mathbf{A}_{\boldsymbol{\theta}_0}^{-1} \mathbf{B}_{\boldsymbol{\theta}_0} \mathbf{A}_{\boldsymbol{\theta}_0}^{-1} \triangleq \text{MCRB}(\boldsymbol{\theta}_0), \tag{5}$$

where

$$\mathbf{C}_p\left(\hat{\boldsymbol{\theta}}(\mathbf{x}), \boldsymbol{\theta}_0\right) \triangleq E_p\left\{\left(\hat{\boldsymbol{\theta}}(\mathbf{x}) - \boldsymbol{\theta}_0\right)\left(\hat{\boldsymbol{\theta}}(\mathbf{x}) - \boldsymbol{\theta}_0\right)^T\right\} \tag{6}$$

is the error covariance matrix of the mismatched estimator $\hat{\boldsymbol{\theta}}(\mathbf{x})$ where the matrices $\mathbf{A}_{\boldsymbol{\theta}_0}$ and $\mathbf{B}_{\boldsymbol{\theta}_0}$ have been defined in eqs. (2) and (3), respectively.

The following comments are in order. The major implication of Theorem 1 is that it is still possible to establish a lower bound on the error covariance matrix of an (MS-unbiased) estimator even if it is derived under a misspecified data model, i.e. it is derived under a pdf $f_X(\mathbf{x}_m | \boldsymbol{\theta})$ that could differ from the true pdf $p_X(\mathbf{x}_m)$ for every value of $\boldsymbol{\theta}$ in the parameter

space $\Theta$. An important question that may arise under a misspecified model framework is which vector in the assumed parameter space $\Theta$ should be used to evaluate the effectiveness of a mismatched estimator, particularly when no "true" parameter vector exists, i.e. $p_X(\mathbf{x}_m) \neq f_X(\mathbf{x}_m | \boldsymbol{\theta})$, for all $\boldsymbol{\theta} \in \Theta$? It is certainly reasonable to use the parameter value that minimizes the "distance", in a given sense, between the assumed misspecified pdf $f_X(\mathbf{x}_m | \boldsymbol{\theta})$ and the true pdf $p_X(\mathbf{x}_m)$. Theorem 1 shows that if, as a measure of the "distance," one uses the KLD and by assuming that the misspecified model $\mathcal{F}$ is regular with respect to the true model $\mathcal{P}$, this parameter vector exists and it is the pseudo-true parameter vector $\boldsymbol{\theta}_0$ defined in eq. (1). Specifically, the MCRB is a lower bound on the error covariance matrix of any MS-unbiased estimator, where the error is defined as the difference between the estimator and the pseudo-true parameter vector. Moreover, if the model $\mathcal{F}$ is correctly specified, then, as said before, $\boldsymbol{\theta}_0 = \bar{\boldsymbol{\theta}}$, such that $p_X(\mathbf{x}_m) = f_X(\mathbf{x}_m | \bar{\boldsymbol{\theta}})$, and $\mathbf{B}_{\boldsymbol{\theta}_0} = \mathbf{B}_{\bar{\boldsymbol{\theta}}} = -\mathbf{A}_{\bar{\boldsymbol{\theta}}}$. Consequently, the inequality in (5) becomes the classical (matched) Cramér-Rao bound inequality for unbiased estimators:

$$E_p\left\{\left(\hat{\boldsymbol{\theta}}(\mathbf{x}) - \bar{\boldsymbol{\theta}}\right)\left(\hat{\boldsymbol{\theta}}(\mathbf{x}) - \bar{\boldsymbol{\theta}}\right)^T\right\} \geq \frac{1}{M}\mathbf{B}_{\bar{\boldsymbol{\theta}}}^{-1} = -\frac{1}{M}\mathbf{A}_{\bar{\boldsymbol{\theta}}}^{-1} \triangleq \text{CRB}(\bar{\boldsymbol{\theta}}). \qquad (7)$$

The second point is how can Theorem 1 be exploited in practice? The MCRB is a generalization of the classical CRB to the misspecified model framework and can play a similar role. Specifically, the MCRB can be used to assess the performance of any mismatched estimator and it plays the same key role as the classical CRB in any feasibility study, but with the added flexibility to assess performance under modelling errors. Consider for example the recurring scenario in which the SP practitioner is aware of the true data pdf $p_X(\mathbf{x}_m)$, but in order to fulfil some operational constraints, the user is forced to derive the required estimator by exploiting a simpler, but misspecified, model. In this scenario, the MCRB in (5) can be directly applied to assess the potential estimation loss due to the mismatch between the assumed and the true models.

This scenario can be extended to the case where the SP practitioner is not completely aware of the functional form of the true data pdf, but the user is still able to infer some of its properties, for example, from empirical data or parameter estimates based on such data. Such knowledge can be used to motivate surrogate models for the true data pdf, which in turn can be exploited to conduct system analysis and performance assessment. To clarify this point, consider the case in which the SP practitioner, in order to derive a simple inference algorithm, decides to assume a Gaussian model to describe the data behaviour. However, thanks to a preliminary data analysis, the user is aware of the fact that the data share a heavy-tailed distribution, e.g., due to the presence of impulsive non-Gaussian noise. Then the user could choose as true data pdf a heavy-tailed distribution, e.g. the *t*-distribution, and consequently, exploit the MCRB to assess how ignoring the heavy-tailed and impulsive nature of the data affects the performance of the estimation algorithm based on a Gaussian model. This explains that, although the chosen "true" pdf (in this example, the *t*-distribution) may not be the exact true data pdf, it can still serve as a useful surrogate for the purposes of system analysis and design by means of the MCRB.

The MCRB can also be used to predict potential weaknesses (i.e. breakdown of the estimation performance) of a system. Suppose one has a system/estimator derived under a certain modelling assumption, but it is of interest for practical reasons to predict how well this system will react in the presence of different "true" input data distributions perhaps characterizing operational scenarios that the system can undergo. Clearly, the MCRB is well-suited to address this task.

Another important question may arise analysing Theorem 1. In order to evaluate the pseudo-true parameter vector $\boldsymbol{\theta}_0$ in eq. (1) and then the MCRB in eq. (5), we need to know the true data pdf $p_X(\mathbf{x}_m)$, since it is required to evaluate the expectation operators. How can we calculate the MCRB in all the practical cases in which we haven't any a priori knowledge of the functional form of $p_X(\mathbf{x}_m)$? An answer to this fundamental question is given in Sect.

4.1, where we show that consistent estimators for both the pseudo-true parameter vector $\boldsymbol{\theta}_0$ and of the MCRB can be derived from the acquired data set.

We conclude this section with two remarks. It is worth mentioning that the proposed MCRB can be easily extended to misspecified estimation problems that require equality constraints. We refer the reader to [For16b] for a comprehensive treatise of this problem. The second remark concerns the possibility to generalize the previously discussed results to the case of *complex* unknown parameter vectors. The extension to the complex fields can be achieved in two equivalent ways. We can always maps a complex parameter vector into a real one simply by stacking its real and the imaginary parts as e.g. in [Ren15] or we could exploit the so-called Wirtinger calculus as discussed in [Ric15] and [For17].

3.4 AN INTERESTING CASE: A LOWER BOUND ON THE MEAN SQUARE ERROR VIA THE MCRB

In this section, we focus on a particular mismatched case that is of great interest in many practical applications. Specifically, we consider the case in which the parameter vector of the assumed model $\mathcal{F}$ is *nested* in the one of the true parametric model $\mathcal{P}$, i.e. the assumed parameter space $\Theta$ is a subspace of the true parameter space $\Omega = \Theta \times \Gamma$, where $\times$ indicates the Cartesian product. Under this restriction, the true parametric model can be expressed as

$$\mathcal{P} = \{p_X \mid p_X(\mathbf{x}_m \mid \boldsymbol{\theta}, \boldsymbol{\gamma}) \text{ is a pdf } \forall (\boldsymbol{\theta}, \boldsymbol{\gamma}) \in \Theta \times \Gamma\}, \tag{8}$$

while the assumed model is $\mathcal{F} = \{f_X \mid f_X(\mathbf{x}_m \mid \boldsymbol{\theta}) \text{ is a pdf } \forall \boldsymbol{\theta} \in \Theta\}$ as before. Note that $f_X(\mathbf{x}_m \mid \boldsymbol{\theta})$ could differ from the true $p_X(\mathbf{x}_m \mid \boldsymbol{\theta}, \boldsymbol{\gamma}) \; \forall \boldsymbol{\theta} \in \Theta$ and $\forall \boldsymbol{\gamma} \in \Gamma$. Moreover, the nested parameter vector assumption includes, as special case, the scenario in which the true parameter space and the assumed one are equal, i.e. $\Omega \equiv \Theta$. This particular case arises, for example, in array processing applications in which both the true and the assumed pdfs of the acquired data vectors can be parameterized by the angles of arrival of a certain number of sources [Ric15]. A practical example of the more general nested model assumption is the estimation of the disturbance covariance matrix in adaptive radar detection [For16a]. In this misspecified estimation problem, both the unknown true data pdf and the assumed one can be parameterized by a scaled version of the covariance matrix and by the disturbance power. Both these applications will be discussed in Sect. 6, while here we focus our attention on the theoretical implications of the condition in (8). The first immediate consequence of (8) is the fact that if the pseudo-true parameter vector $\boldsymbol{\theta}_0$ and the true parameter sub-vector $\bar{\boldsymbol{\theta}}$ belong to the same parameter space $\Theta$, then the difference vector $\mathbf{r} \triangleq \bar{\boldsymbol{\theta}} - \boldsymbol{\theta}_0$ is well defined, but in general is different from a zero-vector. As shown in ([For16a], Sect. II.D) or in ([Ric15], eq. 70), using $\mathbf{r}$, a bound on the *Mean Square Error* (MSE) of the estimate of the true parameter vector $\bar{\boldsymbol{\theta}}$ under model misspecification can be easily established as:

$$\mathrm{MSE}_p\left(\hat{\boldsymbol{\theta}}(\mathbf{x}), \bar{\boldsymbol{\theta}}\right) \triangleq E_p\left\{\left(\hat{\boldsymbol{\theta}}(\mathbf{x}) - \bar{\boldsymbol{\theta}}\right)\left(\hat{\boldsymbol{\theta}}(\mathbf{x}) - \bar{\boldsymbol{\theta}}\right)^T\right\}$$
$$= \mathbf{C}_p\left(\hat{\boldsymbol{\theta}}(\mathbf{x}), \boldsymbol{\theta}_0\right) + \mathbf{r}\mathbf{r}^T \geq \frac{1}{M}\mathbf{A}_{\boldsymbol{\theta}_0}^{-1}\mathbf{B}_{\boldsymbol{\theta}_0}\mathbf{A}_{\boldsymbol{\theta}_0}^{-1} + \mathbf{r}\mathbf{r}^T \triangleq \mathrm{LB}\left(\bar{\boldsymbol{\theta}}\right). \quad (9)$$

Note that here the $\mathrm{LB}(\bar{\boldsymbol{\theta}}) = \mathrm{MCRB}(\boldsymbol{\theta}_0) + \mathbf{r}\mathbf{r}^T$ is considered as a function of the true parameter vector $\bar{\boldsymbol{\theta}}$. A simple example that clarifies the role of the inequality (9) as lower bound on the MSE is reported in Section 4.2.

## 4. THE MISMATCHED MAXIMUM LIKELIHOOD (MML) ESTIMATOR

The aim of this section is to present the second milestone of the estimation theory under model misspecification: the Mismatched Maximum Likelihood (MML) estimator. As discussed in the Introduction, the theoretical framework supporting the existence and the convergence properties of the MML estimator was developed by Huber [Hub67] and later by White [Whi82]. Here, our goal is to summarize their main findings from an SP standpoint. As detailed in Section 2, assume we have a set $\mathbf{x} = \{\mathbf{x}_m\}_{m=1}^M$ of $M$ i.i.d. measurement vectors distributed according to a true, but unknown or inaccessible, pdf $p_X(\mathbf{x}_m)$. So, the *log-likelihood function* for the data $\mathbf{x}$ under a generally misspecified parametric pdf $f_X(\mathbf{x}_m | \boldsymbol{\theta}) \in \mathcal{F}$ is given by $l_M(\boldsymbol{\theta}) \triangleq M^{-1}\sum_{m=1}^{M} \ln f_X(\mathbf{x}_m | \boldsymbol{\theta})$. Following the classical definition, the MML estimate is the vector that maximizes the (misspecifed) log-likelihood function:

$$\hat{\boldsymbol{\theta}}_{MML}(\mathbf{x}) \triangleq \arg\max_{\boldsymbol{\theta} \in \Theta} l_M(\boldsymbol{\theta}) = \arg\max_{\boldsymbol{\theta} \in \Theta} \sum_{m=1}^{M} \ln f_X(\mathbf{x}_m | \boldsymbol{\theta}), \tag{10}$$

where $\mathbf{x}_m \sim p_X(\mathbf{x}_m)$. The definition of the MML estimator given in eq. (10) is clear and self-explanatory. Moreover, it is consistent with the classical "matched" ML estimator. The main question is what is the convergence point of $\hat{\boldsymbol{\theta}}_{MML}(\mathbf{x})$? As proved in [Hub67] and [Whi82], under suitable regularity conditions, the MML estimator converges (*almost surely*, *a.s.*) to the pseudo-true parameter vector $\boldsymbol{\theta}_0$ defined in eq. (1). This is a desirable result, since it shows that the MML estimator converges to the parameter vector that minimizes the distance, in the KLD sense, between the misspecified and the true pdfs. In addition, Huber and White investigated the asymptotic behaviour of the MML estimator and their valuable findings can be summarized in the following theorem.

*Theorem 2 ([Hub67], [Whi82])*: Under suitable regularity conditions, it can be shown that:

$$\hat{\boldsymbol{\theta}}_{MML}(\mathbf{x}) \xrightarrow[M \to \infty]{a.s.} \boldsymbol{\theta}_0. \tag{11}$$

Moreover,

$$\sqrt{M}\left(\hat{\boldsymbol{\theta}}_{MML}(\mathbf{x}) - \boldsymbol{\theta}_0\right) \underset{M \to \infty}{\overset{d.}{\sim}} \mathcal{N}\left(\mathbf{0}, \mathbf{C}_{\boldsymbol{\theta}_0}\right), \tag{12}$$

where $\underset{M \to \infty}{\overset{d.}{\sim}}$ indicates the convergence in distribution and $\mathbf{C}_{\boldsymbol{\theta}_0} \triangleq \mathbf{A}_{\boldsymbol{\theta}_0}^{-1} \mathbf{B}_{\boldsymbol{\theta}_0} \mathbf{A}_{\boldsymbol{\theta}_0}^{-1}$, where the matrices $\mathbf{A}_{\boldsymbol{\theta}_0}$ and $\mathbf{B}_{\boldsymbol{\theta}_0}$ have been defined in eqs. (2) and (3), respectively. Matrix $\mathbf{C}_{\boldsymbol{\theta}_0}$ is sometimes referred to as Huber's "sandwich covariance."

Two comments are in order:

1. The MML estimator is asymptotically MS-unbiased and its asymptotic error covariance is equal to the MCRB, i.e. it is an efficient estimator w.r.t. the MCRB. The analogy with the classical matched ML estimator is completely transparent. In particular, if the model $\mathcal{F}$ is correctly specified, i.e. there exists a parameter vector $\bar{\boldsymbol{\theta}} \in \Theta$ such that $p_X(\mathbf{x}_m) = f_X(\mathbf{x}_m | \bar{\boldsymbol{\theta}})$, then $\hat{\boldsymbol{\theta}}_{MML}(\mathbf{x}) \xrightarrow[M \to \infty]{a.s.} \bar{\boldsymbol{\theta}}$ with an asymptotic error covariance matrix given by the classical CRB, which is the inverse of the FIM $\mathbf{B}_{\bar{\boldsymbol{\theta}}} = -\mathbf{A}_{\bar{\boldsymbol{\theta}}}$.

2. Theorem 2 represents a very useful result for practical applications. In fact, it tells us that, when we do not have any a-priori information about the true data model, the ML estimator derived under a possibly misspecified model, is still a reasonable choice among other MS-unbiased mismatched estimators, since it converges to the parameter vector that minimizes the KLD between the true and the assumed model and it has the lowest possible error covariance (at least asymptotically).

4.1 A CONSISTENT SAMPLE ESTIMATE OF THE MCRB

In this section, we go back to an issue raised before, i.e. the calculation of the MCRB when the true model is completely unknown. In fact, from eq. (5), to obtain a closed form expression of the MCRB, we need to evaluate analytically $\boldsymbol{\theta}_0$, $\mathbf{A}_{\boldsymbol{\theta}_0}$, and $\mathbf{B}_{\boldsymbol{\theta}_0}$. As shown in eqs. (1), (2), and (3), these quantities involve the evaluation of the expectation operator taken w.r.t. the true pdf $p_X(\mathbf{x}_m)$. If $p_X(\mathbf{x}_m)$ is completely unknown, we will not be able to evaluate these expectations in closed form, but, as an alternative, we could obtain sample estimates of them. More formally, we define the matrices [Whi82]:

$$[\mathbf{A}_M(\boldsymbol{\theta})]_{ij} \triangleq M^{-1} \sum_{m=1}^{M} \frac{\partial^2 \ln f_X(\mathbf{x}_m | \boldsymbol{\theta})}{\partial \theta_i \partial \theta_j}, \tag{13}$$

$$[\mathbf{B}_M(\boldsymbol{\theta})]_{ij} \triangleq M^{-1} \sum_{m=1}^{M} \frac{\partial \ln f_X(\mathbf{x}_m | \boldsymbol{\theta})}{\partial \theta_i} \cdot \frac{\partial \ln f_X(\mathbf{x}_m | \boldsymbol{\theta})}{\partial \theta_j}, \tag{14}$$

$$\mathbf{C}_M(\boldsymbol{\theta}) \triangleq [\mathbf{A}_M(\boldsymbol{\theta})]^{-1} \mathbf{B}_M(\boldsymbol{\theta}) [\mathbf{A}_M(\boldsymbol{\theta})]^{-1}. \tag{15}$$

Remarkably, it can be shown (see the proof in [Whi82, Theo 3.2]) that:

$$\mathbf{C}_M(\hat{\boldsymbol{\theta}}_{MML}) \xrightarrow[M \to \infty]{a.s.} \mathbf{C}_{\boldsymbol{\theta}_0} = \text{MCRB}(\boldsymbol{\theta}_0). \tag{16}$$

In other words, eq. (16) assures us that we can obtain a strongly consistent estimate of the MCRB by evaluating the sample counterpart of $\mathbf{A}_{\boldsymbol{\theta}_0}$ and $\mathbf{B}_{\boldsymbol{\theta}_0}$, i.e. $\mathbf{A}_M(\boldsymbol{\theta})$ and $\mathbf{B}_M(\boldsymbol{\theta})$, at the value of the MML estimator. This result has strong practical implications, since it provides an estimate of the MCRB when we do not have any prior knowledge of the true pdf $p_X(\mathbf{x}_m)$. Hence, it widens areas of applicability of the MCRB. This of course requires the data to be stationary over some reasonable period of time to allow sufficient averaging (as is required in numerous SP applications). This result can also be used to design statistical tests to detect model misspecification [Whi82], [Whi96, p. 218].

4.2 AN EXAMPLE: VARIANCE ESTIMATION

We now describe an illustrative example with the aim of clarifying the use and the derivation of the MCRB. Building upon the examples discussed in [For16a], we investigate here the problem of estimating the variance of a Gaussian-distributed dataset under the misspecification of the mean value. Let $\mathbf{x} = \{x_m\}_{m=1}^M$ be a set of $M$ i.i.d. univariate data sampled from a Gaussian pdf with mean value $\bar{\mu}$ and variance $\bar{\sigma}^2$, i.e. $p_X(x_m) \equiv \mathcal{N}(\bar{\mu}, \bar{\sigma}^2) \in \mathcal{P}$ with $\bar{\mu} \neq 0$. Due to perhaps an imperfect knowledge about the data generation process, the user assumes a zero-mean parametric Gaussian model $\mathcal{F} = \{f_X \mid f_X(x_m \mid \theta) = \mathcal{N}(0, \theta) \ \forall \theta \in \Theta \subseteq \mathbb{R}^+\}$, i.e. the user misspecifies the mean value. Note that, as long as $\bar{\mu} \neq 0$, the true but unknown pdf $p_X(x_m)$ does not belong to the assumed model $\mathcal{F}$. Moreover, the reader can easily recognize this mismatched scenario as a simple instance of the particular case discussed in Section 3.4. In fact, it is immediate to verify that the parameter space $\Theta \subseteq \mathbb{R}^+$ that characterizes the assumed model is a subset of the true parameter space, that is $[\bar{\mu}, \bar{\sigma}^2] \in \Omega = \mathbb{R}_0 \times \Theta$, where $\mathbb{R}_0$ indicates the set of all the real numbers excluding 0.

According to the theory presented in Sect. 3, we first have to check if the assumed model $\mathcal{F}$ is regular w.r.t. $\mathcal{P}$ or, in other words, we have to prove the existence of the pseudo-true parameter $\theta_0$ (Assumption A1) and the non-singularity of the matrix $\mathbf{A}_{\theta_0}$ defined in (2) (Assumption A2). Note that, for the problem at hand, $\mathbf{A}_{\theta_0}$ is a scalar quantity, so we have to prove that $A_{\theta_0} \neq 0$. The pseudo-true parameter $\theta_0$ is defined in (1). Following [Cov06], the KLD can be expressed as:

$$D(p_X \| f_{X|\theta}) = \frac{\bar{\mu}^2}{2\theta} + \frac{1}{2}\left(\frac{\bar{\sigma}^2}{\theta} - 1 - \ln\frac{\bar{\sigma}^2}{\theta}\right). \tag{17}$$

The minimum is obtained for $\theta_0 = \bar{\sigma}^2 + \bar{\mu}^2$, which according to (1), represents the pseudo-true parameter. Since the pseudo-true parameter exists and is unique, Assumption A1 is satisfied. We can now check Assumption A2. To this end, from (2), $A_{\theta_0}$ can be evaluated as:

$$A_{\theta_0} \triangleq E_p\left\{\frac{\partial^2 \ln f_X(x_m|\theta)}{\partial \theta^2}\bigg|_{\theta=\theta_0}\right\} = \frac{1}{2\theta_0^2} - \frac{1}{\theta_0^3}E_p\{x_m^2\} = -\frac{1}{2\theta_0^2}, \tag{18}$$

yielding a denominator different from zero since $\bar{\sigma}^2 \in \mathbb{R}^+$; consequently, Assumption A2 is verified as well. Now we can evaluate the MCRB in (5) for the estimation problem at hand. First, the scalar $B_{\theta_0}$ can be easily evaluated from (3) as:

$$B_{\theta_0} \triangleq E_p\left\{\left(\frac{\partial \ln f_X(x_m|\theta)}{\partial \theta}\right)^2\bigg|_{\theta=\theta_0}\right\} = \frac{\theta_0^2 + E_p\{x_m^4\} - 2\theta_0 E_p\{x_m^2\}}{4\theta_0^4} = \frac{2\bar{\sigma}^4 + 4\bar{\sigma}^2\bar{\mu}^2}{4\theta_0^4}. \tag{19}$$

Finally, from (5), we get:

$$\text{MCRB}(\theta_0) = \frac{2\bar{\sigma}^4}{M} + \frac{4\bar{\sigma}^2\bar{\mu}^2}{M}. \tag{20}$$

It can be noted that the MCRB in (20) is always greater than the classical CRB given by $\text{CRB}(\bar{\sigma}^2) = 2\bar{\sigma}^4/M$ and they are equal only in the case of perfect model specification, i.e. when the true mean is equal to the assumed mean, i.e. $\bar{\mu} = 0$.

Since, as said before, this misspecified scenario belongs to the particular class of nested parametric models, discussed in Section 3.4, we can also rewrite the MCRB in eq. (20) as function of the true variance $\bar{\sigma}^2$. This can be easily done by introducing the (scalar) $r \triangleq \bar{\sigma}^2 - \theta_0 = -\bar{\mu}^2$ and consequently, according to eq. (9), by evaluating the $\text{LB}(\bar{\sigma}^2)$ as:

$$\text{LB}(\bar{\sigma}^2) = \frac{2\bar{\sigma}^4}{M} + \frac{4\bar{\sigma}^2\bar{\mu}^2}{M} + \bar{\mu}^4. \tag{21}$$

After having established a lower bound on the MSE, we now investigate the properties of the MML estimator for the estimation problem at hand. In particular, we can say that the MML estimator is not consistent since, from (11), it converges to $\theta_0$, which is different from the true variance $\bar{\sigma}^2$. More formally, we have that:

$$\hat{\theta}_{MML} \triangleq \hat{\theta}_{MML}(\mathbf{x}) = M^{-1}\sum_{m=1}^{M} x_m^2 \xrightarrow[M \to \infty]{a.s.} \theta_0 = \bar{\sigma}^2 + \bar{\mu}^2 \neq \bar{\sigma}^2. \tag{22}$$

However, according to (4), the MML estimator is MS-unbiased since:

$$E_p\{\hat{\theta}_{MML}\} = E_p\left\{M^{-1}\sum_{m=1}^{M} x_m^2\right\} = \bar{\sigma}^2 + \bar{\mu}^2 = \theta_0. \tag{23}$$

Hence, according to Theorem 1, its error covariance w.r.t. $\theta_0$, i.e. $C_p(\hat{\theta}_{MML}, \theta_0)$ is lower bounded by the MCRB in (20). Fig. 1 shows the error covariance of the MML estimator, the $\text{MCRB}(\theta_0)$ and the sample estimate of $\text{MCRB}(\theta_0)$ obtained according to (13)-(15). As we can see, $\text{MCRB}(\theta_0)$ is a tight bound for the error variance of the MML estimator and the sample $\text{MCRB}(\theta_0)$ accurately predicts it. Due to the particular nested structure of the true and assumed parameter spaces of this example, we can also evaluate the MSE of the MML estimator w.r.t. the true variance, i.e. $\text{MSE}_p(\hat{\theta}_{MML}, \bar{\sigma}^2)$ and the related $\text{LB}(\bar{\sigma}^2)$ obtained as shown in (9).

In Fig. 2, we report the MSE of the MML estimator, the $\text{LB}(\bar{\sigma}^2)$ and the classical CRB on the estimation of the variance, $\text{CRB}(\bar{\sigma}^2)$, as function of the value of the true mean value $\bar{\mu}$. As expected from (9), $\text{LB}(\bar{\sigma}^2)$ is a tight bound for the MSE of the MML estimator. Finally, it can be noted that the $\text{LB}(\bar{\sigma}^2)$ is equal to the $\text{CRB}(\bar{\sigma}^2)$ only when $\bar{\mu}=0$, i.e. when the assumed mean value is equal to the true one.

4.3 ANOTHER EXAMPLE: POWER ESTIMATION IN CORRELATED DATA

Another example that clarifies the theory concerns the estimation of the statistical power of a set of zero mean Gaussian vectors. Let $\mathbf{x} = \{\mathbf{x}_m\}_{m=1}^{M}$ be a set of $M$ i.i.d. real $N$-dimensional random vectors sampled from a multivariate Gaussian pdf with zero mean value and covariance matrix given by $\mathbf{M} = \bar{\sigma}^2 \mathbf{\Sigma}$, i.e. $p_X(\mathbf{x}_m) \equiv \mathcal{N}(\mathbf{0}, \bar{\sigma}^2 \mathbf{\Sigma}) \in \mathcal{P}$ where $\bar{\sigma}^2$ is the statistical power and $\mathbf{\Sigma}$ is a symmetric, positive definite matrix whose trace is equal to $N$, i.e. $\text{tr}(\mathbf{\Sigma}) = N$. For simplicity, we assume that $[\mathbf{\Sigma}]_{ij} = \rho^{|i-j|}$, $i,j = 1,\ldots,N$ where $|\rho| < 1$ is the one-lag correlation coefficient (this is the typical correlation matrix of an AR(1) process). Suppose now that the user is not aware of the data correlation structure and decide to assume the following parametric Gaussian model: $\mathcal{F} = \{f_X \mid f_X(\mathbf{x}_m \mid \theta) = \mathcal{N}(\mathbf{0}, \theta \mathbf{I}_N) \; \forall \theta \in \mathbb{R}^+\}$, where $\mathbf{I}_N$ is the identity matrix of dimension $N$. Note that, as long as $\rho \neq 0$, the true pdf $p_X(\mathbf{x}_m)$ does not belong to the assumed model $\mathcal{F}$. We will proceed exactly as in the previous example by checking the assumptions A1 and A2, and then by evaluating the MML estimator and the relative MCRB.

To evaluate the pseudo-true parameter $\theta_0$ we need to find the minimum of the KLD between the true and the assumed model. Following again [Cov06], the KLD between $\mathcal{N}(\mathbf{0}, \bar{\sigma}^2 \mathbf{\Sigma})$ and $\mathcal{N}(\mathbf{0}, \theta \mathbf{I}_N)$ is given by:

$$D(p_X \Vert f_{X|\theta}) = \frac{1}{2} \left[ \text{tr}(\theta^{-1} \bar{\sigma}^2 \mathbf{\Sigma}) - N + \ln\theta - \ln\det(\bar{\sigma}^2 \mathbf{\Sigma}) \right]. \tag{24}$$

Keeping in mind that $\text{tr}(\mathbf{\Sigma}) = N$, it is immediate to verify that the minimum is given by $\theta_0 = \bar{\sigma}^2$, i.e. the pseudo-true parameter is equal to the true power. After some basic calculus, the terms $A_{\theta_0}$ and $B_{\theta_0}$ are obtained as:

$$A_{\theta_0} \triangleq E_p \left\{ \frac{\partial^2 \ln f_X(\mathbf{x}_m | \theta)}{\partial \theta^2} \bigg|_{\theta=\theta_0} \right\} = \frac{N}{2\theta_0^2} - \frac{1}{\theta_0^3} E_p \left\{ \mathbf{x}_m^T \mathbf{x}_m \right\} = -\frac{N}{2\bar{\sigma}^4} , \qquad (25)$$

$$B_{\theta_0} \triangleq E_p \left\{ \left( \frac{\partial \ln f_X(\mathbf{x}_m | \theta)}{\partial \theta} \right)^2 \bigg|_{\theta=\theta_0} \right\} = \frac{N\theta_0^2 + E_p\{(\mathbf{x}_m^T \mathbf{x}_m)^2\} - 2N\theta_0 E_p\{\mathbf{x}_m^T \mathbf{x}_m\}}{4\theta_0^4} = \frac{\mathrm{tr}(\mathbf{\Sigma}^2)}{2\bar{\sigma}^4} . \qquad (26)$$

Finally, from (5), we get:

$$\mathrm{MCRB}(\theta_0) = \mathrm{MCRB}(\bar{\sigma}^2) = \frac{2\bar{\sigma}^4}{MN^2} \mathrm{tr}(\mathbf{\Sigma}^2). \qquad (27)$$

The CRB for the estimation of the statistical power of the true model can be easily obtained as $\mathrm{CRB}(\bar{\sigma}^2) = 2\bar{\sigma}^4/MN$. As expected, the CRB is always greater that the MCRB on $\bar{\sigma}^2$ and they are equal if and only if $\mathbf{\Sigma} = \mathbf{I}$, i.e. when the model is correctly specified. We can go on to investigate the properties of the MML estimator. Unlike the example in Sect 4.1, the MML estimator of the statistical power is consistent since, from (11), it converges to $\theta_0$ that is equal to the true power $\bar{\sigma}^2$:

$$\hat{\theta}_{MML}(\mathbf{x}) = (MN)^{-1} \sum_{m=1}^{M} \mathbf{x}_m^T \mathbf{x}_m \xrightarrow[M \to \infty]{a.s.} \theta_0 = \bar{\sigma}^2 . \qquad (28)$$

Moreover, the MML estimator is MS-unbiased since:

$$E_p\{\hat{\theta}_{MML}(\mathbf{x})\} = (MN)^{-1} \sum_{m=1}^{M} E_p\{\mathbf{x}_m^T \mathbf{x}_m\} = N^{-1} \bar{\sigma}^2 \mathrm{tr}(\Sigma) = \bar{\sigma}^2 = \theta_0 , \qquad (29)$$

And then, according to Theorem 1, its MSE is lower bounded by the MCRB in (20). Fig. 3 shows the MSE of the MML estimator, the MCRB, the sample estimate of the MCRB, and the CRB as function of the one-lag coefficient $\rho$. The MCRB is a tight bound for the MSE of the MML estimator and the sample MCRB accurately predicts it. Finally, we note that the MCRB is equal to the CRB only when $\rho = 0$, i.e. when $\mathbf{\Sigma} = \mathbf{I}$.

**5. GENERALIZATION TO THE BAYESIAN SETTING**

The Bayesian philosophy adopts the notion that one has some prior knowledge (belief or perhaps a guess) about the values a desired parameter will assume before an experiment. Once data are observed, then one can update that prior knowledge based on the information provided by the data measurements. Thus, the Bayesian framework is designed to allow prior knowledge to influence the estimation process in an optimal fashion. Specifically, within a Bayesian framework, estimation of the parameter vector $\boldsymbol{\theta}$ is derived from the joint pdf $f_{X,\boldsymbol{\theta}}(\mathbf{x},\boldsymbol{\theta})$ instead of solely the conditional (non-Bayesian) pdf $f_{X|\boldsymbol{\theta}}(\mathbf{x}|\boldsymbol{\theta})$. From basic probability theory, the joint density can be expressed as $f_{X,\boldsymbol{\theta}}(\mathbf{x},\boldsymbol{\theta}) = f_{\boldsymbol{\theta}|X}(\boldsymbol{\theta}|\mathbf{x})f_X(\mathbf{x})$ where clearly the posterior density $f_{\boldsymbol{\theta}|X}(\boldsymbol{\theta}|\mathbf{x})$ summarizes all the information needed to make any inference on $\boldsymbol{\theta}$ based on the data $\mathbf{x} = \{\mathbf{x}_m\}_{m=1}^{M}$. The joint density can likewise be related to the conditional density that models the parameter's influence on data measurements, i.e. $f_{X,\boldsymbol{\theta}}(\mathbf{x},\boldsymbol{\theta}) = f_{X|\boldsymbol{\theta}}(\mathbf{x}|\boldsymbol{\theta})f_{\boldsymbol{\theta}}(\boldsymbol{\theta})$. Prior knowledge about parameter vector $\boldsymbol{\theta}$ is reflected in the prior pdf $f_{\boldsymbol{\theta}}(\boldsymbol{\theta})$. When there is no prior knowledge, all outcomes for the parameter vector can be assumed equally likely. Such a non-informative prior often leads to results consistent with standard non-Bayesian approaches, i.e. yields algorithms and bounds that rely primarily on $f_{X|\boldsymbol{\theta}}(\mathbf{x}|\boldsymbol{\theta})$. Thus, the Bayesian framework in a sense can be considered a generalization of the non-Bayesian framework [Van07], [Van13], [Leh98].

When the model is perfectly specified, the optimal Bayesian estimator under cost metrics such as the *squared error* and the *uniform cost*, depends primarily on the posterior distribution $f_{\boldsymbol{\theta}|X}(\boldsymbol{\theta}|\mathbf{x})$. Indeed, the squared error cost is minimized by the conditional mean estimator $\hat{\boldsymbol{\theta}}_{MSE}(\mathbf{x}) = E_{f_{\boldsymbol{\theta}|X}}\{\boldsymbol{\theta}|\mathbf{x}\}$ and the uniform cost is minimized by the maximum a posteriori (MAP) estimator $\hat{\boldsymbol{\theta}}_{MAP}(\mathbf{x}) = \arg\max_{\boldsymbol{\theta}} f_{\boldsymbol{\theta}|X}(\boldsymbol{\theta}|\mathbf{x})$ [Van13], [Leh98]. Under perfect model specification the asymptotic properties of Bayes estimators and the posterior distribution have been investigated extensively. It is known that under suitable conditions, as the number of data samples increases the Bayes estimator tends to become independent of the prior distribution ([Leh98], Ch. 4). Thus, the influence of the prior distribution on posterior inferences decreases and asymptotic behaviour similar to the non-Bayesian ML estimator emerges. Indeed, strong consistency, efficiency, and normality properties of Bayes estimators have been established for a large class of prior distributions [Str81]. This asymptotic behaviour has some intuitive appeal since the prior represents a statistical summary of one's best guess (prior to an actual experiment) of the likelihood the desired parameter will assume any particular value. As actual data measurements become available, however, it makes sense that one will eventually abandon the guidance provided by the prior pdf in light of the valuable information carried by the data measurements obtained from the actual experiment. This phenomenon is well established and has been observed in signal processing applications. When the prior $f_{\boldsymbol{\theta}}(\boldsymbol{\theta})$ is incorrect but the model $f_{X|\boldsymbol{\theta}}(\mathbf{x}|\boldsymbol{\theta})$ is correct then it is possible that a significantly larger number of data observations (or higher signal-to-noise ratios) may be required before the Bayes estimator becomes independent of the influence of the incorrect prior ([Kan13] p. 4737).

Misspecification within a Bayesian framework explores the possibility that the assumed joint pdf $f_{X,\boldsymbol{\theta}}(\mathbf{x},\boldsymbol{\theta})$ may be incorrect. This, of course, includes the prior pdf $f_{\boldsymbol{\theta}}(\boldsymbol{\theta})$ as well as the model $f_{X|\boldsymbol{\theta}}(\mathbf{x}|\boldsymbol{\theta})$. Under model misspecification, the asymptotic properties of the posterior distribution also have been investigated extensively. The following discussion attempts to summarize some key results on this topic, although no claims are made here that the summary is complete or exhaustive. The goal here is to identify results perhaps of interest to the signal processing community in the authors' viewpoint. The first discussion to follow will focus on published results that detail the asymptotic behaviour and properties of the Bayesian posterior distribution under model misspecification, i.e. the asymptotic behaviour of $f_{\boldsymbol{\theta}|X}(\boldsymbol{\theta}|\mathbf{x})$ as the amount of data increases. These results can be considered the Bayesian counterparts in spirit of the contributions of Huber [Hub67] and White [Whi82] that detail ML estimator performance under misspecification, as discussed before. Secondly, a discussion of results on mis-specified Bayesian bounds is given. As this remains a relatively new area of research, there appear to be very few published results on this topic. Hence, a brief discussion of some of the issues involved is given.

5.1 BAYESIAN ESTIMATION UNDER MISSPECIFIED MODELS

Since Bayes estimators are derived from the posterior density $f_{\boldsymbol{\theta}|X}(\boldsymbol{\theta}|\mathbf{x})$, considering its asymptotic behaviour yields insights into the convergence properties of the associated estimators. Berk [Ber66] was the first one to investigate the asymptotic behaviour of the posterior distribution under misspecification as the number of data observations becomes arbitrarily large. Specifically, consider a set of i.i.d. data measurements $\mathbf{x} = \{\mathbf{x}_m\}_{m=1}^{M}$ according to joint pdf $p_X(\mathbf{x}) = \prod_{m=1}^{M} p_X(\mathbf{x}_m)$. Let the *assumed* pdf of $\mathbf{x}$ be $f_X(\mathbf{x}|\boldsymbol{\theta}) = \prod_{m=1}^{M} f_X(\mathbf{x}_m|\boldsymbol{\theta})$ and the assumed prior be $f_{\boldsymbol{\theta}}(\boldsymbol{\theta})$. Define the set $\Theta_A$ such that

$$\Theta_A \triangleq \left\{ \boldsymbol{\theta} \in \Theta : \arg\min_{\boldsymbol{\theta} \in \Theta} \left\{ -E_p \left\{ \ln f_X(\mathbf{x}|\boldsymbol{\theta}) \right\} \right\} \right\}. \tag{30}$$

For a large class of unimodal and well-behaved distributions, the set $\Theta_A$ consists of a single unique point, i.e. $\Theta_A = \{\boldsymbol{\theta}_0\}$, but clearly the definition allows for the possibility that this set contains more than one point. It is also noteworthy (see also eq. (1)) that the set $\Theta_A$ is simply the set of all points/vectors $\boldsymbol{\theta} \in \Theta$ that minimizes the KLD $D(p_X \| f_{X|\boldsymbol{\theta}})$ between the true and assumed distributions. Berk noted this relation to the KLD in [Ber66], i.e. prior to the Akaike [Aka72] reference to Huber's work [Hub67]. In particular, Berk proved that, if $\Theta_A = \{\boldsymbol{\theta}_0\}$, i.e. it consists of a single unique point $\boldsymbol{\theta}_0$, then the following convergence in distribution holds:

$$f_{\boldsymbol{\theta}|X}(\boldsymbol{\theta}|\mathbf{x}) \triangleq f_{\boldsymbol{\theta}|X}(\boldsymbol{\theta}|\mathbf{x}_1, \mathbf{x}_2, \ldots, \mathbf{x}_M) \xrightarrow[M \to \infty]{d.} \delta(\boldsymbol{\theta} - \boldsymbol{\theta}_0), \tag{31}$$

where $\delta(\mathbf{a}) = \delta(a_1)\delta(a_2)\cdots\delta(a_d)$ and $\delta(a)$ is a Dirac delta function.

From (31), one can presume that $\boldsymbol{\theta}_0$ is the counterpart for the misspecified Bayesian estimation framework of the pseudo-true parameter vector introduced in (1). This conjecture is validated by the fundamental results of Bunke and Milaud [Bun98] that provide strong consistency arguments for a class of *Mismatched* (or *pseudo*) *Bayesian (MB) estimators*. Specifically, let $L(\cdot,\cdot)$ be a nonnegative, real valued loss function such that $L(\boldsymbol{\theta},\boldsymbol{\theta})=0$. A familiar example of this type of function is the one leading to the MSE between a given estimate $\hat{\boldsymbol{\theta}}$ and a given vector $\boldsymbol{\theta}$, i.e. $L_{MSE}(\hat{\boldsymbol{\theta}},\boldsymbol{\theta})=(\hat{\boldsymbol{\theta}}-\boldsymbol{\theta})^T(\hat{\boldsymbol{\theta}}-\boldsymbol{\theta})$. Consider now the class of (possibly mismatched) Bayesian estimates defined as:

$$\hat{\boldsymbol{\theta}}_{MB}(\mathbf{x}) \triangleq \arg\min_{\vartheta \in \Theta} E_{f_{\boldsymbol{\theta}|X}}\left\{L(\vartheta,\boldsymbol{\theta})\right\} = \arg\min_{\vartheta \in \Theta} \int_{\Theta} L(\vartheta,\boldsymbol{\theta}) f_{\boldsymbol{\theta}|X}(\boldsymbol{\theta}\,|\,\mathbf{x}) d\boldsymbol{\theta}. \tag{32}$$

Bunke and Milaud [Bun98] investigated the asymptotic behaviour of the class of estimators in (32) and their results can be recast as follows.

*Theorem 3 ([Bun98])*: Under certain regularity conditions (see A1-A11 in [Bun98]) and provided that $\Theta_A = \{\boldsymbol{\theta}_0\}$, it can be shown that:

$$\hat{\boldsymbol{\theta}}_{MB}(\mathbf{x}) \underset{M \to \infty}{\overset{a.s.}{\to}} \boldsymbol{\theta}_0. \tag{33}$$

Moreover,

$$\sqrt{M}\left(\hat{\boldsymbol{\theta}}_{MB}(\mathbf{x}) - \boldsymbol{\theta}_0\right) \underset{M \to \infty}{\overset{d.}{\sim}} \mathcal{N}\left(\mathbf{0},\boldsymbol{\Lambda}_{\boldsymbol{\theta}_0}\right), \tag{34}$$

where

$$\boldsymbol{\Lambda}_{\boldsymbol{\theta}_0} \triangleq \bar{\mathbf{L}}_2^{-1}\bar{\mathbf{L}}_1 \mathbf{A}_{\boldsymbol{\theta}_0}^{-1} \mathbf{B}_{\boldsymbol{\theta}_0} \mathbf{A}_{\boldsymbol{\theta}_0}^{-1} \left(\bar{\mathbf{L}}_2^{-1}\bar{\mathbf{L}}_1\right)^T, \tag{35}$$

$$\left[\bar{\mathbf{L}}_1\right]_{i,j} = \left.\frac{\partial^2 L(\boldsymbol{\alpha},\boldsymbol{\beta})}{\partial \alpha_i \partial \beta_j}\right|_{\substack{\boldsymbol{\alpha}=\boldsymbol{\theta}_0 \\ \boldsymbol{\beta}=\boldsymbol{\theta}_0}}, \quad \left[\bar{\mathbf{L}}_2\right]_{i,j} = \left.\frac{\partial^2 L(\boldsymbol{\alpha},\boldsymbol{\theta}_0)}{\partial \alpha_i \partial \alpha_j}\right|_{\boldsymbol{\alpha}=\boldsymbol{\theta}_0}, \tag{36}$$

and the matrices $\mathbf{A}_{\boldsymbol{\theta}_0}$ and $\mathbf{B}_{\boldsymbol{\theta}_0}$ have been defined in eqs. (2) and (3), respectively.

Two comments are in order:

1. The similarity between the results given in Theorem 1 for the MML estimator and the ones given in Theorem 2 for the MB estimator is now clear: under model misspecification (and under suitable regularity conditions), both the MML and the MB estimators converge almost surely to the point $\theta_0$ that minimizes the KLD between the true and the assumed distributions. Moreover they are both asymptotically normal distributed with covariance matrices that are related to the matrices $\mathbf{A}_{\theta_0}$ and $\mathbf{B}_{\theta_0}$.

2. If, in (32), the squared error loss function $L_{MSE}(\boldsymbol{\alpha}, \boldsymbol{\beta})$ is used, then $\bar{\mathbf{L}}_1 = -\bar{\mathbf{L}}_2 = 2\mathbf{I}$ and, consequently, the asymptotic covariance matrices of the MB estimator and the MML estimator are the same, i.e. $\boldsymbol{\Lambda}_{\theta_0} = \mathbf{C}_{\theta_0} = \mathbf{A}_{\theta_0}^{-1} \mathbf{B}_{\theta_0} \mathbf{A}_{\theta_0}^{-1}$.

While identifying key results from [Bun98] and [Ber66] in this article, reference has been made to several assumptions (see e.g. A1-A11 in [Bun98]) whose details were omitted here. While important (in particular, the uniqueness of the KLD minimizer is critical in Theorem 3), inclusion of these details would unnecessarily clutter the discussion. However, the regularity conditions described by [Bun98] characterize a wide spectrum of problems relevant to the signal processing community.

To conclude, the results discussed in this section are based on a parametric model $f_X(\mathbf{x}|\boldsymbol{\theta})$ for the data. It is worth mentioning that a similar convergence persists in the non-parametric case. Specifically, Kleijn and Van der Vaart [Kle06] address convergence properties of the posterior distribution in the nonparametric case as well as the rate of convergence.

5.2 BAYESIAN BOUNDS UNDER MISSPECIFIED MODELS

As sketched in the Introduction, when the model is correctly specified, a wide family of Bayesian bounds can be derived from the covariance inequality [Van07]. As well detailed in [Van07] and [Ren08], this family includes the Bayesian Cramér-Rao Bound, the Bayesian Bhattacharyya Bound, the Bobrovsky-Zakai Bound, and the Weiss-Weinstein Bound, among others. Establishing Bayesian bounds under model misspecification appears to have received very limited attention and represents an area of open research. The only results on the topic to the authors' knowledge are given in [Kan15] and [Ric16]. The approach taken therein differs from the classic approach adopted in [Van07] with some loss in generality. In fact, the Bayesian bounds obtained in [Kan15] and [Ric16] attempt to build on the non-Bayesian results in [Ric15]. Specifically, it is required that the true conditional pdf $p_{X|\theta}(\mathbf{x}|\boldsymbol{\theta})$ and the assumed model $f_{X|\theta}(\mathbf{x}|\boldsymbol{\theta})$ share the same parameter space $\Theta$, thus any misspecification is exclusively due to the functional form of the assumed distribution. This is essentially the particular case discussed in the non-Bayesian context in sub-section 3.4, and the bound that we are going to derive has a form similar to the non-Bayesian bound in (9).

Let the conditional mean of the estimator be $E_{p_{X|\theta}}\{\hat{\boldsymbol{\theta}}(\mathbf{x})\} = \boldsymbol{\mu}(\boldsymbol{\theta})$ and define the error vector and the bias vector as $\boldsymbol{\zeta}(\mathbf{x},\boldsymbol{\theta}) \triangleq \hat{\boldsymbol{\theta}}(\mathbf{x}) - \boldsymbol{\theta}$ and $\mathbf{r}(\boldsymbol{\theta}) \triangleq \boldsymbol{\mu}(\boldsymbol{\theta}) - \boldsymbol{\theta}$, respectively. As in (9), the total MSE is given by the sum of the covariance and squared bias. Thus, by use of the covariance inequality [Van07] a lower bound on MSE under model misspecification is given by:

$$\text{MSE}_{p_{X,\theta}}\left(\hat{\boldsymbol{\theta}}(\mathbf{x})\right) \triangleq E_{p_{X,\theta}}\left\{\boldsymbol{\zeta}\boldsymbol{\zeta}^T\right\} \geq \frac{1}{M} E_{p_{X,\theta}}\left\{\boldsymbol{\zeta}\boldsymbol{\eta}^T\right\} E_{p_{X,\theta}}^{-1}\left\{\boldsymbol{\eta}\boldsymbol{\eta}^T\right\} E_{p_{X,\theta}}\left\{\boldsymbol{\eta}\boldsymbol{\zeta}^T\right\} + E_{p_\theta}\left\{\mathbf{r}\mathbf{r}^T\right\}, \quad (37)$$

where we dropped the dependences on $\mathbf{x}$ and $\boldsymbol{\theta}$ for notation simplicity. The vector function $\boldsymbol{\eta}(\mathbf{x},\boldsymbol{\theta})$ represents the *score function* [Van07] and a judicious choice of it leads to tight bounds. In [Kan15] and [Ric16], the following score function is considered with the aim of obtaining a bound for the Bayes MAP estimator and ML estimator in mind:

$$\boldsymbol{\eta}(\mathbf{x},\boldsymbol{\theta}) = \nabla_{\boldsymbol{\theta}} \ln f_{X|\theta}(\mathbf{x}|\boldsymbol{\theta}) - E_{p_{X|\theta}}\left\{\nabla_{\boldsymbol{\theta}} \ln f_{X|\theta}(\mathbf{x}|\boldsymbol{\theta})\right\}. \quad (38)$$

This score function is the same as the one used for the MCRB in [Ric15] and it leads to a version of the misspecified Bayesian CRB (MBCRB). To provide a sketch of this fact, we define the following two matrices based on the conditional expectation: $E_{p_{X|\theta}}\left\{\boldsymbol{\eta}\boldsymbol{\zeta}^T\right\} = \boldsymbol{\Xi}(\boldsymbol{\theta})$ and $E_{p_{X|\theta}}\left\{\boldsymbol{\eta}\boldsymbol{\eta}^T\right\} = \mathbf{J}(\boldsymbol{\theta})$. Closed-form expressions can be found in [Ric15] for the case where both the true and the assumed conditional distributions are complex Gaussian, for example. The resulting lower bound on MSE follows from (37) and is given by:

$$\text{MSE}_{p_{X,\theta}}\left(\hat{\boldsymbol{\theta}}(\mathbf{x})\right) \geq \frac{1}{M} E_{p_\theta}\left\{\boldsymbol{\Xi}^T(\boldsymbol{\theta})\right\} E_{p_\theta}^{-1}\left\{\mathbf{J}(\boldsymbol{\theta})\right\} E_{p_\theta}\left\{\boldsymbol{\Xi}(\boldsymbol{\theta})\right\} + E_{p_\theta}\left\{\mathbf{rr}^T\right\}. \tag{39}$$

The class of estimators to which the above MBCRB applies is that with mean and estimator-score function correlation satisfying respectively

$$E_{p_{X,\theta}}\left\{\hat{\boldsymbol{\theta}}(\mathbf{x})\right\} = E_{p_\theta}\left\{\boldsymbol{\mu}(\boldsymbol{\theta})\right\}, \quad E_{p_{X,\theta}}\left\{\boldsymbol{\eta}(\mathbf{x},\boldsymbol{\theta})[\hat{\boldsymbol{\theta}}(\mathbf{x}) - \boldsymbol{\mu}(\boldsymbol{\theta})]^T\right\} = E_{p_\theta}\left\{\boldsymbol{\Xi}(\boldsymbol{\theta})\right\}. \tag{40}$$

These constraints follow from the covariance inequality ([Ric15], Sect. III-C) and the choice of score function. This limits the applicability of the bound in contrast to bounds obtained when the model is perfectly specified. Thus, an obvious area of future effort is the development of Bayesian bounds under misspecified models with fewer constraints and broader applicability. To conclude, we note that an example demonstrating the applicability of this Bayesian Bound to Direction of Arrival (DOA) estimation for sparse arrays is given in [Kan15].

## 6. EXAMPLES OF APPLICATIONS

In this section, we describe some examples related to the problems of DOA estimation and data covariance/scatter matrix estimation. These problems are relevant in many array processing and adaptive radar applications.

### 6.1 DOA ESTIMATION UNDER MODEL MISSPECIFICATION

The estimation of the DOAs of plane wave signals by means of an array of sensors has been the core research area within the array signal processing community for years [Van02]. The fundamental prerequisite for any DOA estimation algorithm is that the positions of the sensors in the array are known exactly, i.e. known geometry. Many authors have investigated the impact of imperfect knowledge of the sensor positions on the DOA estimation perform-ance, or of the misscalibration of the array itself (see e.g. [Fri90] and [Van02], just to name two of them). Other authors have proposed hybrid or modified CRBs with the aim to predict the MSE of the DOA estimators in the presence of the position uncertainties ([Roc87], [Par08]). The goal of this section is to show that the misspecified estimation framework presented in this paper is a valuable and general tool to deal with modelling errors in the array manifold. The application of the MCRB and the MML estimator to the DOA estimation problem has been recently investigated in [Ric15] for Uniform Linear Arrays (ULAs) and in [Ren15] for MIMO radar systems.

Following [Ric15], consider a ULA of $N$ sensors and a single plane wave signal impinging on the array from a conic angle $\bar{\theta}$. Moreover, suppose that, due to an array misscalibration, the true position vector $\mathbf{p}_n$ of the $n^{th}$ sensor, defined in a three-dimensional Cartesian coordinate frame, is known up to an error term modelled as a zero-mean, Gaussian random vector, i.e. $\mathbf{e}_n \sim \mathcal{N}(0, \sigma_e^2 \mathbf{I}_3)$. Then, the received data can be expressed as $x_n = s[\mathbf{d}(\bar{\theta})]_n + [\mathbf{c}]_n$, where $[\mathbf{d}(\bar{\theta})]_n = \exp(j\mathbf{k}_{\bar{\theta}}^T(\mathbf{p}_n + \mathbf{e}_n))$ is the $n^{th}$ element of the true (perturbed) steering vector and $\mathbf{k}_{\bar{\theta}} = (2\pi/\lambda)\mathbf{u}(\bar{\theta})$, where $\mathbf{u}(\bar{\theta})$ is a unit vector pointing at the direction defined by $\bar{\theta}$ and $\lambda$ is the wavelength of the transmitted signal. Moreover, $s$ is an unknown deterministic complex scalar that accounts for the transmitted power, the source scattering characteristics and the two-way path loss while $\mathbf{c} = \mathbf{n} + \mathbf{j}$ is the disturbance noise term composed of white Gaussian noise $\mathbf{n}$ and of interference signal (or jammer) $\mathbf{j}$. Given particular realizations of the position errors $\mathbf{e}_n$, the clutter vector is usually modelled as a zero-mean complex Gaussian random vector $\mathbf{c} \sim \mathcal{N}(0, \mathbf{I}_N + \sigma_j^2 \mathbf{d}(\theta_j)\mathbf{d}(\theta_j)^H)$ where $\sigma_j^2$ and $\theta_j$ represent the power and the DOA of the jamming signal. The DOA estimation problem is clearly the estimation of $\bar{\theta}$ given the complex data vector $\mathbf{x}$. Since in practice, it is quite impossible to be aware of the particular realizations of the position error vectors $\mathbf{e}_n$, the user may decide to derive a DOA estimator starting from the *nominal* steering vector $\mathbf{v}(\bar{\theta})$, whose components are $[\mathbf{v}(\bar{\theta})]_n = \exp(j\mathbf{k}_{\bar{\theta}}^T \mathbf{p}_n)$, i.e. the user neglects the sensor position errors. The true (unknown) data model is given by the pdf $p_X(\mathbf{x}) = \mathcal{N}(\bar{s}\mathbf{d}(\bar{\theta}), \mathbf{I}_N + \sigma_j^2 \mathbf{d}(\theta_j)\mathbf{d}(\theta_j)^H) \in \mathcal{P}$, while the assumed parametric model is:

$$\mathcal{F} = \{f_X \mid f_X(\mathbf{x}|s,\theta) = \mathcal{N}(s\mathbf{v}(\theta), \mathbf{I}_N + \sigma_j^2 \mathbf{v}(\theta_j)\mathbf{v}(\theta_j)^H), \forall s \in \mathbb{C}, \theta \in [0, 2\pi)\}. \quad (41)$$

It must be noted that the true pdf $p_X(\mathbf{x})$ does not belong to $\mathcal{F}$, or in other words, the assumed parametric pdf $f_X(\mathbf{x}|s,\theta)$ differs from $p_X(\mathbf{x})$ for every value of $\theta \in [0, 2\pi)$. This is because, even if both the true and the assumed pdfs are complex Gaussian, by neglecting the position errors in the assumed steering vector, we are choosing the wrong parameterization for the mean value and the covariance matrix of the assumed Gaussian model. The question that naturally arises is how large is the performance loss due to this model mismatch? The MCRB presented in Sect. 3 answers this question. We omit the details of the calculation of the MCRB and the derivation of the joint MML estimator of the DOA and of the scalar *s*. We refer the readers to [Ric15]. However, in order to provide some insights about this mismatched estimation problem, Fig. 4 illustrates the matched CRB in the estimation of $\bar{\theta}$, i.e. the CRB on the DOA estimation evaluated by considering the true data pdf $p_X(\mathbf{x})$, the MCRB and the MSE of the MML estimator obtained from the assumed and misspecified pdf $f_X(\mathbf{x}|s,\theta)$. Fig. 4 plots the square roots of the bounds and of the MSE (RMSE) in units of beamwidths as a function of element level SNR. The MCRB accurately predicts performance of the MML estimator. If the system goal is a 10-to-1 beamsplit ratio, i.e. -10dB RMSE in beamwidths, then this could be accomplished with an SNR of 9.28dB when the model is perfectly known, but not knowing precisely the true sensor positions requires an additional ~10dB of *SNR* to achieve the same goal ($\mathrm{MCRB} \cong -10\,\mathrm{dB}$ for $SNR \cong 19.4\,\mathrm{dB}$). On the other hand, if the system receives an $SNR \cong 9.3\,\mathrm{dB}$, then minimum achievable beamsplit ratio in the presence of array errors is 3-to-1, i.e. the $\mathrm{MCRB} \cong -5\,\mathrm{dB}$ RMSE in beamwitdths. This information can be quite valuable in determining where to focus efforts for improve system performance.

## 6.2 SCATTER MATRIX ESTIMATION UNDER MODEL MISSPECIFICATION

Another widely encountered inference problem is the estimation of the correlation structure, i.e. the scatter or covariance matrix, of a dataset. Estimation of the covariance/scatter matrix is a central component of a wide variety of SP applications [Oll12]: adaptive detection and DOA estimation in array processing, Principal Component Analysis (PCA), signal separation, interference cancellation and the portfolio optimization in finance, just to name a few. Even if the data may come from disparate applications, they usually share a non-Gaussian, heavy-tailed statistical nature, as discussed e.g. in [Zou12]. Estimating the covariance matrix of a set of non-Gaussian data, however, is not a trivial task. In fact, non-Gaussian distribution characterization typically requires additional parameters that have to be jointly estimated along with the scatter matrix. Think for example to the (complex) $t$-distribution that has been widely adopted as a suitable and flexible model able to characterize the non-Gaussian, heavy-tailed data behaviour [Lan89], [San12], [Oll12]. A complex, zero-mean, random vector $\mathbf{x}_m \in \mathbb{C}^N$ is said to be $t$-distributed if its pdf can be expressed as:

$$p_X(\mathbf{x}_m | \bar{\mathbf{\Sigma}}, \lambda, \eta) \triangleq \frac{1}{\pi^N |\bar{\mathbf{\Sigma}}|} \frac{\Gamma(N+\lambda)}{\Gamma(\lambda)} \left(\frac{\lambda}{\eta}\right)^\lambda \left(\frac{\lambda}{\eta} + \mathbf{x}_m^H \bar{\mathbf{\Sigma}}^{-1} \mathbf{x}_m\right)^{-(N+\lambda)}, \quad \text{tr}(\bar{\mathbf{\Sigma}}) = N, \qquad (42)$$

where $\Gamma(\cdot)$ indicates the Gamma function while $\lambda$ and $\eta$ are the so-called shape and scale parameters and $\bar{\mathbf{\Sigma}}$ is the scatter matrix. This multidimentional pdf is obtained by assuming that vector $\mathbf{x}_m$ follows the compound-Gaussian model with Gaussian speckle and inverse-Gamma distributed texture [San12]. For proper identifiability, a constraint on $\bar{\mathbf{\Sigma}}$, e.g. $\text{tr}(\bar{\mathbf{\Sigma}}) = N$, needs to be imposed. The complex $t$-distribution has tails heavier than the Gaussian for every $\lambda \in (0, \infty)$, and it becomes the complex Gaussian distribution for $\lambda \to \infty$. As can be clearly seen from (42), in order to perform some inference on a $t$-distributed dataset, we have to jointly estimate the shape and scale parameters along with the scatter matrix. Unfortunately, as pointed out in [Lan89], a joint ML estimator of these three quantities presents convergence and even existence issues. Moreover, as discussed in Sect. 3.3, the $t$-distribution may be only an ap-proximation of the true heavy-tailed data model. To overcome these problems, the SP practitioner has fundamentally two choices: *i*) to apply some robust covariance matrix estimator (see [Oll12] and [Zou12] for further details) or *ii*) to assume a simpler, but generally misspecified, model for characterizing the data, gaining the possibility to derive a closed-form estimator at the cost of a loss in the estimation performance [For16a], [For16c]. If option *ii*) is adopted, the most reasonable choice for the simplified data model is the Complex Gaussian distribution:

$$f_X(\mathbf{x}_m | \boldsymbol{\theta}) \triangleq f_X(\mathbf{x}_m | \boldsymbol{\Sigma}, \sigma^2) = \frac{1}{(\pi\sigma^2)^N |\boldsymbol{\Sigma}|} \exp\left(-\frac{\mathbf{x}_m^H \boldsymbol{\Sigma}^{-1} \mathbf{x}_m}{\sigma^2}\right), \quad \mathrm{tr}(\boldsymbol{\Sigma}) = N. \qquad (43)$$

In fact, the joint (constrained) MML estimator of the scatter matrix and of the data power can be derived as:

$$\hat{\boldsymbol{\Sigma}}_{CMML} = \frac{N}{\sum_{m=1}^{M} \mathbf{x}_m^H \mathbf{x}_m} \sum_{m=1}^{M} \mathbf{x}_m \mathbf{x}_m^H, \quad \hat{\sigma}^2_{CMML} = \frac{1}{NM} \sum_{m=1}^{M} \mathbf{x}_m^H \hat{\boldsymbol{\Sigma}}_{CMML}^{-1} \mathbf{x}_m. \qquad (44)$$

Two comments are in order:

1. It can be shown that $\hat{\boldsymbol{\Sigma}}_{CMML}$ converges to the true scatter matrix, i.e. $\hat{\boldsymbol{\Sigma}}_{CMML} \xrightarrow[M \to \infty]{a.s.} \bar{\boldsymbol{\Sigma}}$, thus it can be successfully applied to estimate it [For16a], [For16c].

2. It is computationally inexpensive and easy to implement which makes the use of $\hat{\boldsymbol{\Sigma}}_{CMML}$ feasible in real-time applications, e.g. in adaptive radar detection.

Along with the knowledge of the convergence point of the MML estimator, it is of great interest to assess the performance loss due to model mismatch. To this purpose, since the Gaussian model is nested in heavy-tailed *t*-distributed model (see Section 3.4), we can evaluate the MCRB for the problem at hand and compare it with the CRB. As an example in Fig. 5 we compare the curves relative to the constrained CRB (CCRB) for the estimation of the scatter matrix under matched conditions (i.e. when the true *t*-distribution is assumed), the constrained

MCRB (CMCRB) [For16b] (i.e. when the misspecified Gaussian model is assumed), and the MSE of the constrained MML estimator of eq. (44) (details of the calculations can be found in [For16c]). The distance between the CCRB and the CMCRB curves provides a measure of the performance loss due to model mismatch. As expected, the loss increases when the shape parameter $\lambda$ goes to zero, i.e. when the data have an extremely heavy-tailed behaviour. On the other hand, when $\lambda \to \infty$, i.e. when the *t*-distribution tends to the Gaussian one, the CCRB and the CMCRB tend to coincide. We note that the constrained MML estimator of the scatter matrix is an efficient estimator w.r.t. the CMCRB, as predicted by the theory in Sect. 4.

## 7. CONCLUDING REMARKS

The objective of this paper is to provide an accessible, and at the same time comprehensive, treatment of the fundamental concepts about Cramér-Rao bounds and efficient estimators in the presence of model misspecification. Every SP practitioner is well aware of the fact that, in almost all practical applications, a certain amount of mismatch between the true and the assumed statistical data models is inevitable. Despite its ubiquity, the assessment of performance bounds under model misspecification appears to have received limited attention from the SP community, while it has been deeply investigated by the statistical community. The first aim of this tutorial paper is to propose to a wide SP audience a comprehensive review of the main contributions to the mismatched estimation theory, both for the deterministic and Bayesian frameworks, with a particular focus on the derivation of CRB under model mismatching. Specifically, we have described how the classical tools of the estimation theory can be generalized to address a mismatched scenario. Firstly, the MCRB has been introduced and the behaviour of the MML estimator investigated. Secondly, results related to the deterministic estimation framework have been extended to the Bayesian one. The existence and the asymptotic properties of a mismatched Bayesian estimator have been discussed. Moreover, some general ideas about the possibility to derive misspecified Bayesian Cramér-Rao Bounds have been provided. In the last part of the paper, we showed how to apply the theoretical findings to two well-known relevant problems: the DOA estimation in array processing and the estimation of the disturbance covariance matrix for adaptive radar detection.

Of course, much work remains to be done. In the following, we try to identify some open problems that could be of great interest for the SP community. A question that naturally arises is whether it is possible to derive a more general class of misspecified bounds. The first step toward this direction has been outlined by Richmond and Horowitz in [Ric15], where a generalization of the theory to the Bhattacharyya bound, to the Barankin bound, and to the Bobrovsky-Mayer-Wolf-Zakai bound has been proposed. Secondly, as discussed in Section 5, a future area of research is the derivation of general Bayesian lower bounds that could be obtained by relaxing or, hopefully, removing the constraints given in (40). Thirdly, a systematic and deep investigation of a general decision theory under model misspecification is required since it could lead great advantages in a huge number of SP applications.

**ACKNOWLEDGMENT.** *The work of Stefano Fortunati has been partially supported by the Air Force Office of Scientific Research under award number FA9550-17-1-0065.*

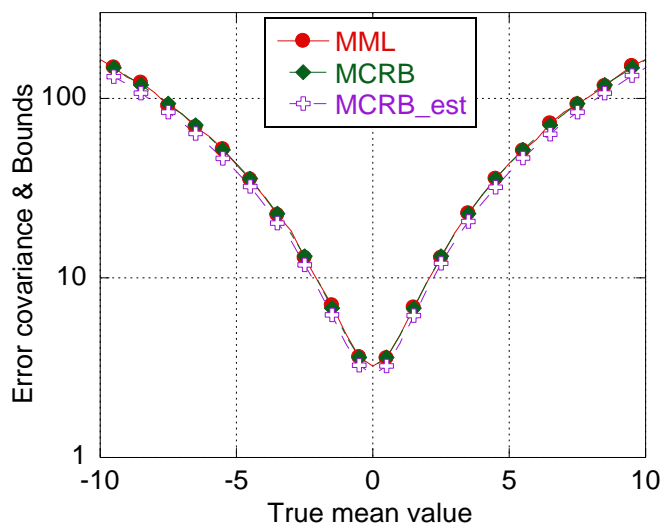

Fig. 1 – Error covariance of the MML estimator, MCRB($\theta_0$) and estimated MCRB($\theta_0$) as function of $\bar{\mu}$. Simulation parameters: $M=10$ and $\bar{\sigma}^2=4$.

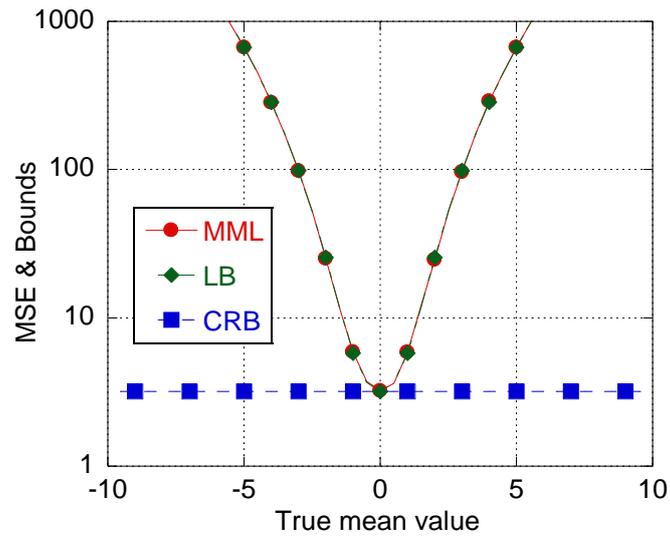

Fig. 2 – MSE of the MML estimator, $\text{LB}(\bar{\sigma}^2)$, and $\text{CRB}(\bar{\sigma}^2)$ as function of $\bar{\mu}$. Simulation parameters: $M=10$ and $\bar{\sigma}^2=4$.

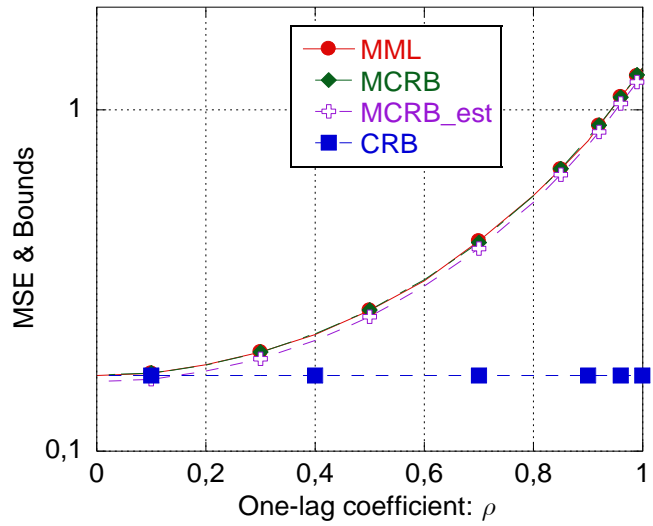

Fig. 3 – MSE of the MML estimator, MCRB, estimated MCRB, and CRB as function of $\rho$. Simulation parameters: $N=8$, $M=3N$ and $\bar{\sigma}^2=4$.

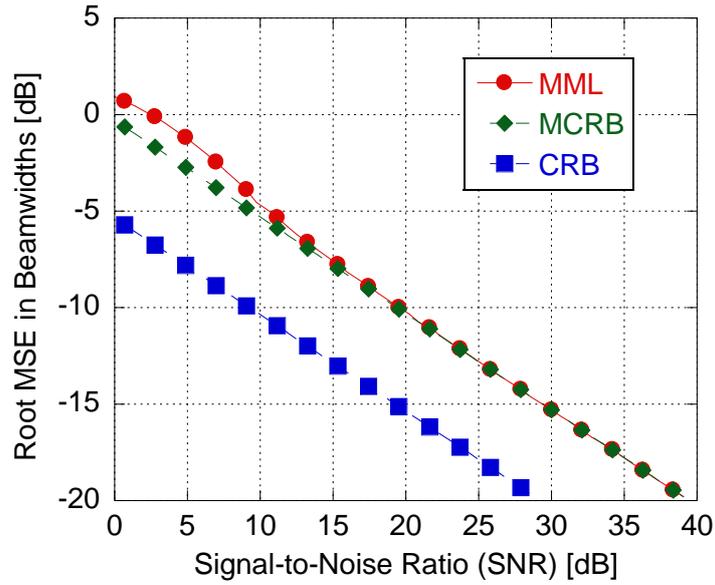

Fig. 4 − MSE of the MML estimator, MCRB, for the DOA estimation problem. Simulation parameters: $M=18$ element ULA, the array position errors of $\sigma_e=0.01\lambda$ of standard deviation, $\theta_t=90°$, $\theta_j=87°$ and $\sigma_J^2=10^3$ (see [Ric15]).

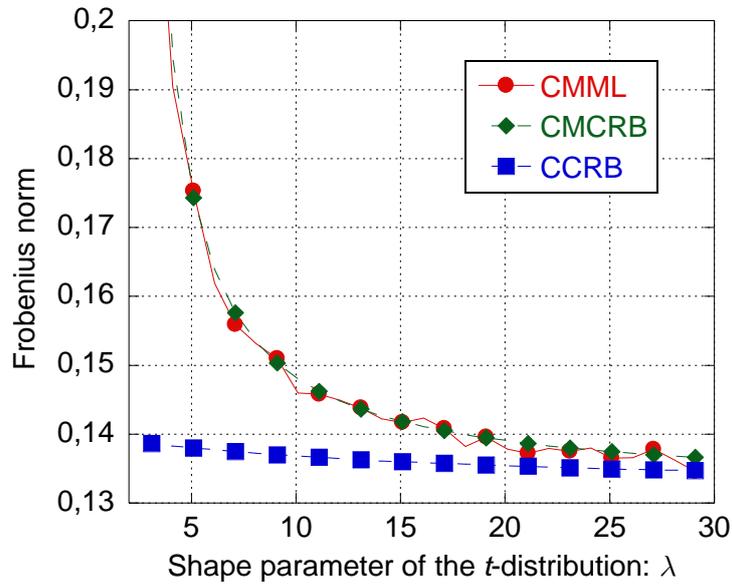

Fig. 5 − Frobenius norms of the MSE matrix of the CMML estimator, CMCRB and CCRB for the scatter matrix estimation problem. Simulation parameters: $N=16$, $M=10N$, the scale parameter of the true $t$-distribution is $\eta=1$.